# Introducing GPGPUs to smartphone-based digital holographic microscope for 3D imaging

YUKI NAGAHAMA*

*Institute of Engineering, Tokyo University of Agriculture and Technology, 2-24-16 Naka-cho, Koganei, Tokyo 184-8588, Japan*
*\*yuki-nagahama@go.tuat.ac.jp*

**Digital holography (DH) enables non-contact, noninvasive 3D imaging of transparent and moving microscopic samples by capturing amplitude and phase information in a single shot. In this work, we present a compact, low-cost, real-time smartphone-based DHM system accelerated by GPUs. The system comprises a 3D-printed optical system using readily available image sensors and lasers, coupled with an Android app for hologram reconstruction, extracting amplitude and phase information. Results show a frame rate improvement of approximately 1.65x compared to a CPU-only system. This inexpensive, compact DHM, combining 3D-printed optics and smartphone-based reconstruction, offers a novel approach compared to existing systems and holds promise for fieldwork and remote diagnostics.**

## 1. Introduction

Digital holography (DH) reconstructs 3D object images from holograms recorded using light waves. It's valuable in bio-imaging due to its single-shot capture of amplitude and phase information, enabling non-contact, noninvasive observation of transparent and moving objects. Digital holographic microscopy (DHM) combines DH with microscopy.

DHM is used to study cell dynamics (division, red blood cell membranes, stem cells) and diagnose diseases like sickle cell anemia [4-7]. Numerous custom DHM systems exist, employing techniques like two-wavelength recording [8, 9, 10], spatial frequency multiplexing [11, 12], cross-reference holographic microscopy [13], low-coherence illumination [14], multimodal imaging [15], Fresnel biprisms [16], and diffractive phase microscopy [17]. However, these systems are typically large and expensive.

Low-cost, portable DHM is desirable for applications like fieldwork and remote diagnostics. Miniaturization efforts have explored integrating smartphones [18-25] as image sensors and computational units. However, smartphone variations require custom interface components, and existing smartphone DHMs often reconstruct holograms on external devices [23, 24, 26] or perform non-real-time reconstruction on the phone itself [25].

Our previous work [27] proposed a DHM system using a USB camera to capture holograms, with real-time reconstruction on an Android smartphone. However, the frame rate (1.92 fps) needs improvement for observing moving objects. This study aims to accelerate processing by leveraging the smartphone's built-in GPU for hologram reconstruction.

## 2. Principle

We used the Gabor-type optical system used in our previous work [27] as the optical system for DHM (Fig. 1). The system (Fig. 1(a), 1(b)) was 3D-printed (101 x 50 x 55 mm, excluding cables) and used a disassembled USB camera (ELECOM UCAM-C980FBBK) as the image sensor. Figure 1(c) shows the system connected to and operating with a smartphone.

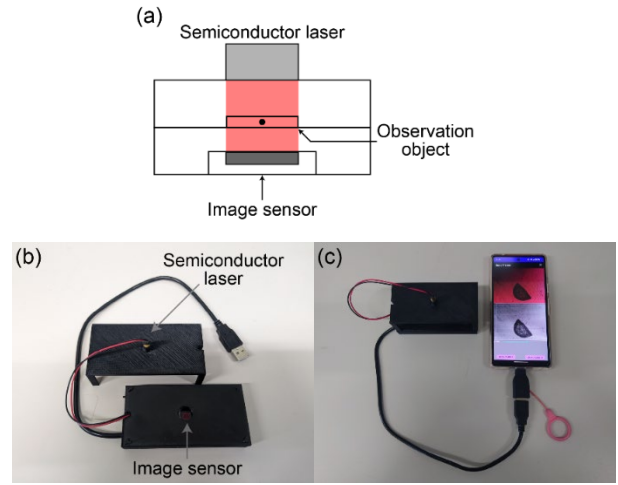

Figure 1. (a) Schematic of the Gabor-type optical system. (b) The optical system of the proposed DHM. (c) The optical system of the proposed DHM in operation.

Smartphones have limited computational power and memory. Therefore, we used band-limited double-step Fresnel diffraction (BL-DSF) [28] to reduce data points and accelerate hologram reconstruction. Diffraction calculations fall into convolution-based (e.g., angular spectrum method - ASM) and Fourier transform-based types. The ASM is shown below:

$$u_2(x_2, y_2) = \text{FFT}^{-1}\left[\text{FFT}[u_1(x_1, y_1)] \exp\left(-2\pi i z \sqrt{1/\lambda^2 - f_x^2 - f_y^2}\right)\right] \quad (1)$$

Where $\lambda$ is the wavelength, $\text{FFT}[\cdot]$ and $\text{FFT}^{-1}[\cdot]$ are the fast Fourier transform and its inverse, $u_1(x_1,y_1)$ and $u_2(x_2,y_2)$ represent the source and destination planes, respectively, $(f_x, f_y)$ are frequency domain coordinates, and $z$ is the propagation distance. Convolution-based diffraction methods like ASM offer the advantage of identical sampling rates between source and destination planes. However, FFT convolution is circular, requiring extension of the input field to perform linear convolution. This requires zero-padding and expanding the source and destination planes to $2N \times 2N$ (where $N$ is the hologram's horizontal and vertical pixel count). Consequently, ASM's memory usage and computational cost scale with $4N^2$ and $4N^2 \log 4N$, respectively, leading to increased resource demands.

To address this, double-step Fresnel diffraction (DSF) [29] was developed. DSF calculates light propagation from the source to the destination plane via a virtual plane $(x_v, y_v)$ using two Fourier transform-based calculations. Because DSF uses Fourier transforms, zero-padding is unnecessary. While most Fourier transform methods alter the sampling rate, DSF allows independent control of source plane sampling rates $p_s$ and destination plane sampling rates $p_d$ by adjusting the distances $z_1$ and $z_2$ to the virtual plane: $p_d = |z_1/z_2| p_s$. BL-DSF further incorporates band-limiting (using a rectangular function) to prevent aliasing. BL-DSF is expressed in Eq. (2).

Where $z_1$ and $z_2$ are the propagation distances from the source to the virtual plane and the virtual to the destination plane, respectively, and $C_{z_2} = \exp\left(\frac{i\pi}{\lambda z_2}(x_2^2 + y_2^2)\right)$. The operator $\text{FFT}^{sgn(z)}$ denotes a forward FFT if $z$ is positive and an inverse FFT if z is negative. The bandwidth restriction area calculation is detailed in [28]. Because BL-DSF is not convolutional, its memory usage and computational cost are proportional to $N^2$ and $N^2 \log N$, respectively.

The Android OS itself provides a function to acquire images from a USB camera on an Android smartphone [30], but whether this function is implemented depends on the device. Therefore, in this study, we used a library called UVCCamera [31] to acquire images from a USB camera on an Android smartphone. In addition, the part that processes holograms captured by the USB camera is implemented using OpenCV [32]. The OpenCV library for Android distributed in [32] does not support OpenCL, a library for parallel computation on GPUs. So, OpenCL is built together with OpenCV source code so that OpenCV functions can be processed using GPUs. This allows GPU programming on C++ using the Android NDK [33].

In the hologram reconstruction calculation, the process of calculating and generating the terms that change the phase (such as $C_{z_2}$ in equation (2)) takes time if done one pixel at a time, so this part was parallelized by writing OpenCL kernel code to process it. Figure 2 shows how programming languages and libraries are used. For convenience, the Android application developed this time is called PocketHoloScope.

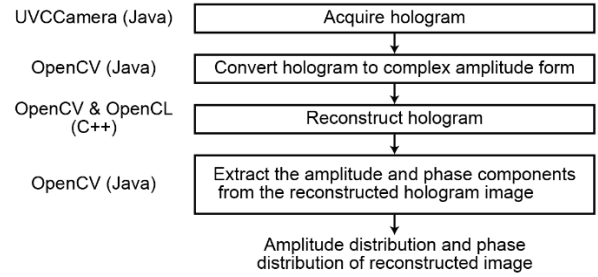

Figure 2. Flow from hologram acquisition to reconstruction and the associated libraries and languages used in PocketHoloScope.

## 2. Experiment

Figure 3 shows a PocketHoloScope screenshot. A seek bar adjusts the hologram reconstruction propagation distance. A button toggles between amplitude and phase display. Pinch gestures control zoom, and a save button saves the reconstructed image. Table 1 lists reconstruction parameters. The USB camera's 3264x2448 sensor resolution was downsampled to 1920x1440 for efficient processing. PocketHoloScope ran on a Google Pixel 9 Pro (Android 15).

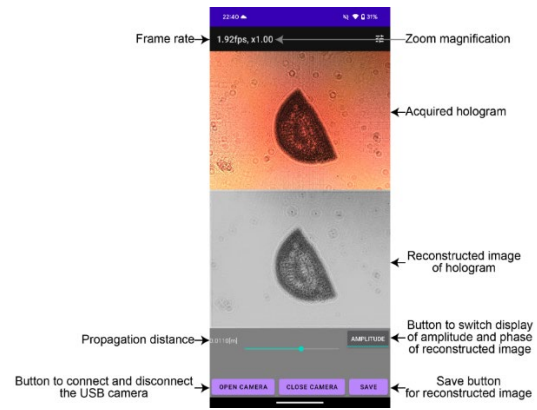

Figure 3. Screenshot of PocketHoloScope

Table 1. Hologram reconstruction computation conditions

| Image sensor resolution | 3,264 × 2,448 pixels |
|---|---|
| Image sensor pixel pitch | 1.47 μm |
| Hologram resolution | 1920 × 1,440 pixels |
| Hologram sampling rate | 2.50 μm |
| Laser wavelength | 650 nm |

Figure 4(a) shows the amplitude components of the reconstructed image when the object of observation is a pine leaf, c.s., and the light propagation distance is focused on the object of observation (0.011 m), using BL-DSF for the

$$u_2(m_2, n_2) = C_{z_2} \text{FFT}^{sgn(z_2)}\left[\exp\left(\frac{i\pi z(x_v^2 + y_v^2)}{\lambda z_1 z_2}\right) \text{Rect}\left(\frac{x_v}{x_v^{\max}}, \frac{y_v}{y_v^{\max}}\right) \text{FFT}^{sgn(z_1)}\left[u_1(m_1, n_1)\exp\left(\frac{i\pi(x_1^2 + y_1^2)}{\lambda z_1}\right)\right]\right] \quad (2)$$

hologram reproduction calculations with CPU alone. The BL-DSF $z_1$ and $z_2$ are adjusted so that the sampling rate of the hologram matches the sampling rate of the reconstructed image. Furthermore, the measured frame rate was found to be 1.75 fps. Figure 4(b) also shows the phase components of the reconstructed image calculated in the same way.

Figure 4(c) also shows the amplitude components of the reconstructed image when calculated with both CPU and GPU under the same conditions as in Figure 4(a). The measured frame rate was found to be 2.89 fps. Figure 4(d) also shows the phase component of the reconstructed image, calculated in the same way.

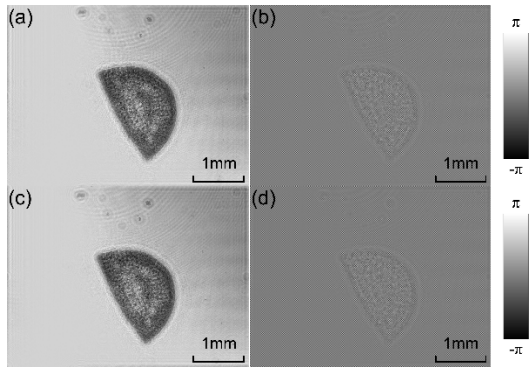

Figure 4. (a) Amplitude component of reconstructed hologram image of pine leaf, c.s. (with CPU). (b) Phase component of reconstructed hologram image of pine leaf, c.s. (with CPU). (c) Amplitude component of reconstructed hologram image of pine leaf, c.s. (with CPU and GPU). (d) Phase component of reconstructed hologram image of pine leaf, c.s. (with CPU and GPU).

Comparing Figures 4(a) and 4(c), and Figures 4(b) and 4(d), respectively, the observed images are almost similar when computed by the CPU alone and by both the CPU and GPU. On the other hand, when comparing the frame rate, the frame rate was 1.75 fps when calculated using only the CPU, and 2.89 fps when calculated using both the CPU and GPU, which is about 1.65 times faster when calculated using the GPU.

### 3. Conclusion

We discuss the frame rate of the proposed system. As mentioned in the "Experiment" section, the DHM system proposed in this study acquired, reconstructed, and displayed holograms at a frame rate of 2.89 fps when using both CPU and GPU and at that of 1.75 fps when using only CPU. This result demonstrates the value of implementing GPGPUs in acquiring, reconstructing, and displaying holograms in real time.

In GPGPUs, the data exchange between the CPU and GPU is often the bottleneck. Therefore, the degree of speedup depends on the target of GPGPU. In the Appendix, we describe how to build a development environment for GPGPU using smartphones in the hope of discovering more suitable computation targets for GPGPU using smartphones.

### 4. Back matter

*5.1 Funding*

Japan Society for the Promotion of Science (21K17760).

*5.2 Disclosures.*

The authors declare no conflicts of interest.

*5.3 Data availability.*

Data underlying the results presented in this paper are not currently publicly available, but may be obtained from the authors upon request.

# Appendix:GPGPU on smartphones using OpenCV and OpenCL

Yuki Nagahama

## Preparation

OpenCV4.10.0 Source code: https://github.com/opencv/opencv/tree/4.10.0

Opencv_contrib 4.10.0 Source code:
https://github.com/opencv/opencv_contrib/tree/4.10.0

CMake3.31.5: https://cmake.org/download/

NDKr27b: available on Android Studio

Zulu-11(jdk) https://www.azul.com/downloads/

MinGW: https://github.com/niXman/mingw-builds-binaries/releases

apache-ant-1.10.15: https://ant.apache.org/bindownload.cgi

・Place OpenCV4.10.0 Source code, Opencv_contrib 4.10.0 Source code, CMake3.31.5, MinGW, apache-ant-1.10.15 all in the "D:/dev" directory.

・I will proceed with the explanation using the username "yuki".

## Build OpenCV

Environment Variable Settings

Manually create a new system environment variable.

```
ANDROID_NDK_HOME = C:¥Users¥yuki¥AppData¥Local¥Android¥Sdk¥ndk¥27.1.12297006
ANDROID_HOME = C:¥Users¥yuki¥AppData¥Local¥Android¥Sdk
JAVA_HOME = C:¥Program Files¥Zulu¥zulu-11
```

## PATH Variables

%JAVA_HOME%\bin

%ANDROID_HOME%\tools

%ANDROID_HOME%\platform-tools

%ANDROID_NDK_HOME%\prebuilt\windows-x86_64\bin

D:\dev\mingw64\bin

D:\dev\cmake-3.31.5-windows-x86_64\bin

D:\dev\apache-ant-1.10.15\bin

## CMAKE Settings

```
CMake 3.31.5 - D:/dev/opencv-4.10.0/build
File  Tools  Options  Help

Where is the source code:     D:/dev/opencv-4.10.0                                          Browse Source...
Preset:                       <custom>
Where to build the binaries:  D:/dev/opencv-4.10.0/build                                    Browse Build...

Search:                                              □ Grouped  ☑ Advanced  ➕ Add Entry  ✖ Remove Entry  Environment...

Name                                     Value
ANDROID_ABI                              arm64-v8a
ANDROID_STL_TYPE                         c++_static
ANDROID_TOOLCHAIN                        clang
ANT_EXECUTABLE                           D:/dev/apache-ant-1.10.15/bin/ant.bat
BLAS_Accelerate_LIBRARY                  BLAS_Accelerate_LIBRARY-NOTFOUND
BLAS_acml_LIBRARY                        BLAS_acml_LIBRARY-NOTFOUND
BLAS_acml_mp_LIBRARY                     BLAS_acml_mp_LIBRARY-NOTFOUND
BLAS_armpl_lp64_LIBRARY                  BLAS_armpl_lp64_LIBRARY-NOTFOUND
BLAS_blas_LIBRARY                        BLAS_blas_LIBRARY-NOTFOUND
BLAS_blastrampoline_5_LIBRARY            BLAS_blastrampoline_5_LIBRARY-NOTFOUND
BLAS_blis_LIBRARY                        BLAS_blis_LIBRARY-NOTFOUND
BLAS_complib_sgimath_LIBRARY             BLAS_complib_sgimath_LIBRARY-NOTFOUND
BLAS_cxml_LIBRARY                        BLAS_cxml_LIBRARY-NOTFOUND
BLAS_dxml_LIBRARY                        BLAS_dxml_LIBRARY-NOTFOUND
BLAS_eml_LIBRARY                         BLAS_eml_LIBRARY-NOTFOUND
BLAS_essl_LIBRARY                        BLAS_essl_LIBRARY-NOTFOUND
BLAS_fjlapack_LIBRARY                    BLAS_fjlapack_LIBRARY-NOTFOUND
BLAS_flexiblas_LIBRARY                   BLAS_flexiblas_LIBRARY-NOTFOUND
BLAS_goto2_LIBRARY                       BLAS_goto2_LIBRARY-NOTFOUND
BLAS_mkl_intel_c_LIBRARY                 BLAS_mkl_intel_c_LIBRARY-NOTFOUND
BLAS_mkl_intel_lp64_LIBRARY              BLAS_mkl_intel_lp64_LIBRARY-NOTFOUND
BLAS_mkl_rt_LIBRARY                      BLAS_mkl_rt_LIBRARY-NOTFOUND
BLAS_openblas_LIBRARY                    BLAS_openblas_LIBRARY-NOTFOUND
BLAS_scs_LIBRARY                         BLAS_scs_LIBRARY-NOTFOUND
BLAS_sgemm_LIBRARY                       BLAS_sgemm_LIBRARY-NOTFOUND

Press Configure to update and display new values in red, then press Generate to generate selected build files.

Configure    Generate    Open Project   Current Generator: MinGW Makefiles

    Other third-party libraries:
      Lapack:                         NO
      Eigen:                          NO
      Custom HAL:                     NO
      Protobuf:                       build (3.19.1)
      Flatbuffers:                    builtin/3rdparty (23.5.9)

    OpenCL:                           YES (SVM NVD3D11)
      Include path:                   D:/dev/opencv-4.10.0/3rdparty/include/opencl/1.2
      Link libraries:                 Dynamic load

    Python (for build):               C:/Users/yuki/AppData/Local/Programs/Python/Python312/python.exe

    Java:
      ant:                            D:/dev/apache-ant-1.10.15/bin/ant.bat (ver 1.10.15)
      Java:                           NO
      JNI:                            C:/Program Files/Zulu/zulu-11/include C:/Program Files/Zulu/zulu-11/include/win32 C:/Program Fi
      Java wrappers:                  YES (ANT)
      Java tests:                     YES

    Install to:                       D:/dev/opencv-4.10.0/build/install
    -----------------------------------------------------------------

Configuring done (8.9s)
Generating done (1.8s)
```

Enter "D:¥dev¥opencv-4.10.0" for Where is the source code and "D:¥dev¥opencv-4.10.0¥build" Where to build the binaries and click 'Configure'.

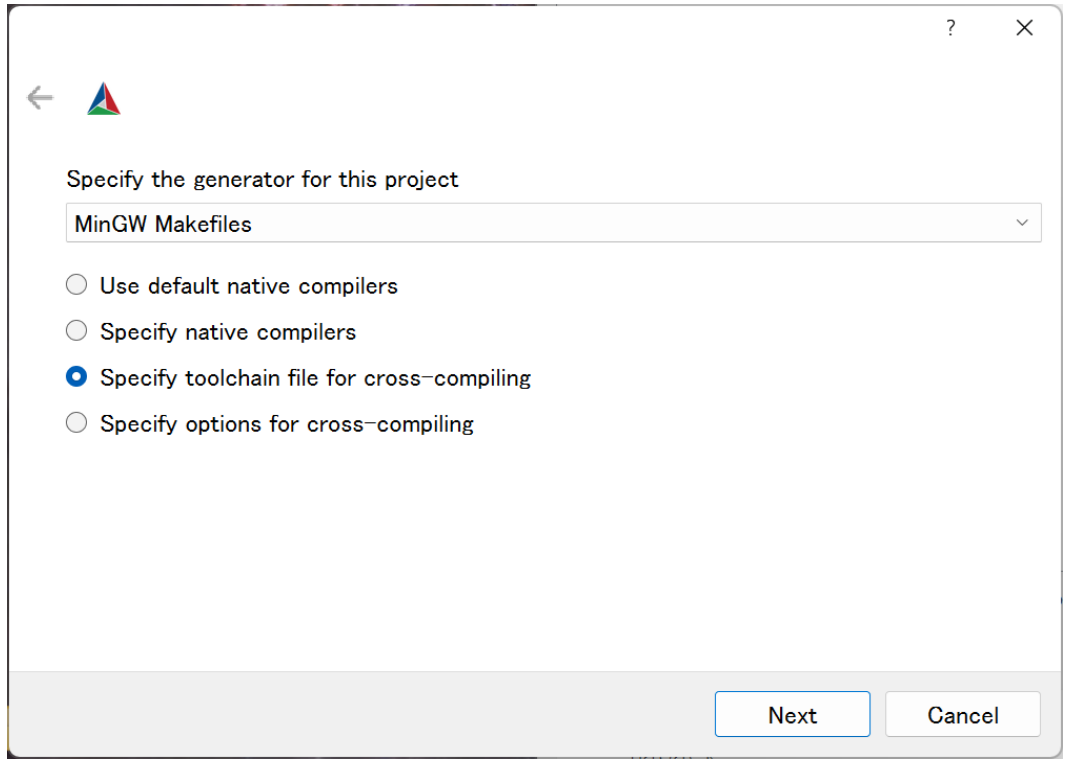

Select "MinGW Makefiles" and "Specify toolchain file for cross-compiling" as shown in the figure and click Next.

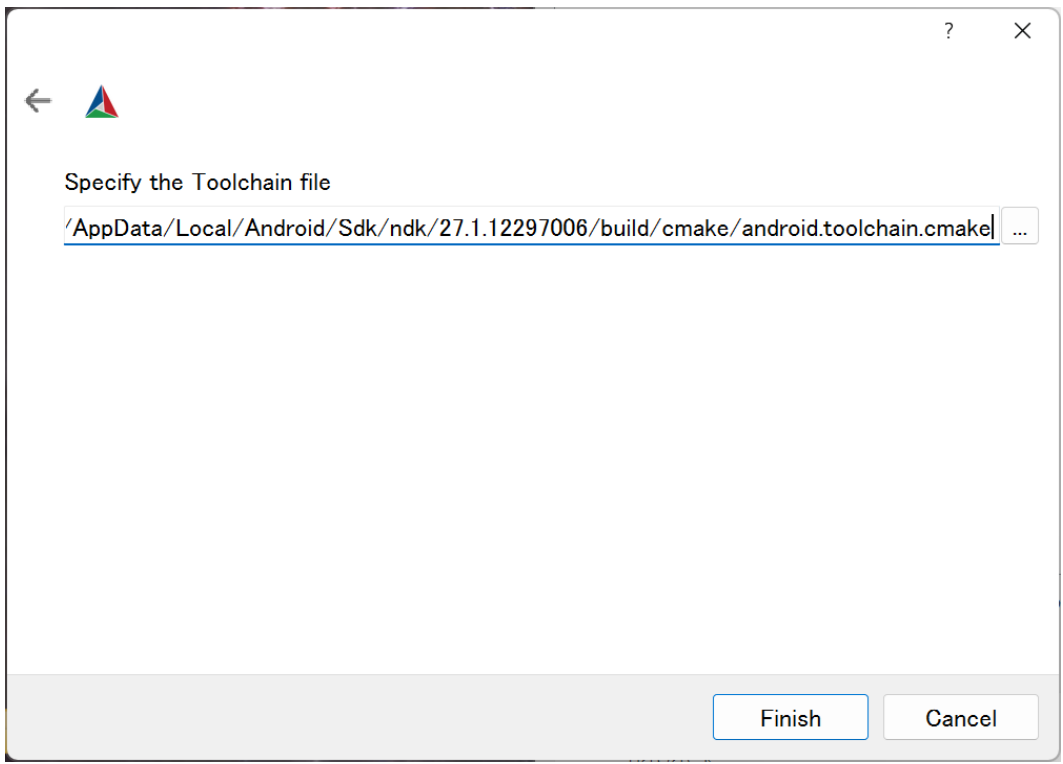

Specify the Toolchain file as

"C:/Users/yuki/AppData/Local/Android/Sdk/ndroid/ndk/27.1.12297006/build/cmake/android.toolchain. cmake" and click 'Finish'.

CMAKE GUI Option Setting

```
ANDROID_ABI:STRING=arm64-v8a
ANDROID_STL_TYPE:STRING=c++_static
ANDROID_TOOLCHAIN:STRING=clang
BUILD_ANDROID_EXAMPLES:BOOL=OFF
BUILD_ANDROID_PROJECTS:BOOL=OFF
BUILD_TESTS:BOOL=OFF
BUILD_PERF_TESTS:BOOL=OFF
OPENCV_ENABLE_NONFREE:BOOL=ON
OPENCV_EXTRA_MODULES_PATH:PATH=D:/dev/opencv_contrib-4.10.0/modules
ANT_EXECUTABLE:PATH=D:/dev/apache-ant-1.10.15/bin
BUILD_opencv_world:BOOL=OFF
WITH_OPENCL=ON
WITH_OPENCL_SVM=ON
OPENCV_DISABLE_FILESYSTEM_SUPPORT=ON
```

Besides, set the following options to compile the dynamic library (.so).

```
BUILD_FAT_JAVA_LIB:BOOL=OFF
BUILD_SHARED_LIBS:BOOL=ON
```

After setting the CMAKE GUI options, click 'Generate'.

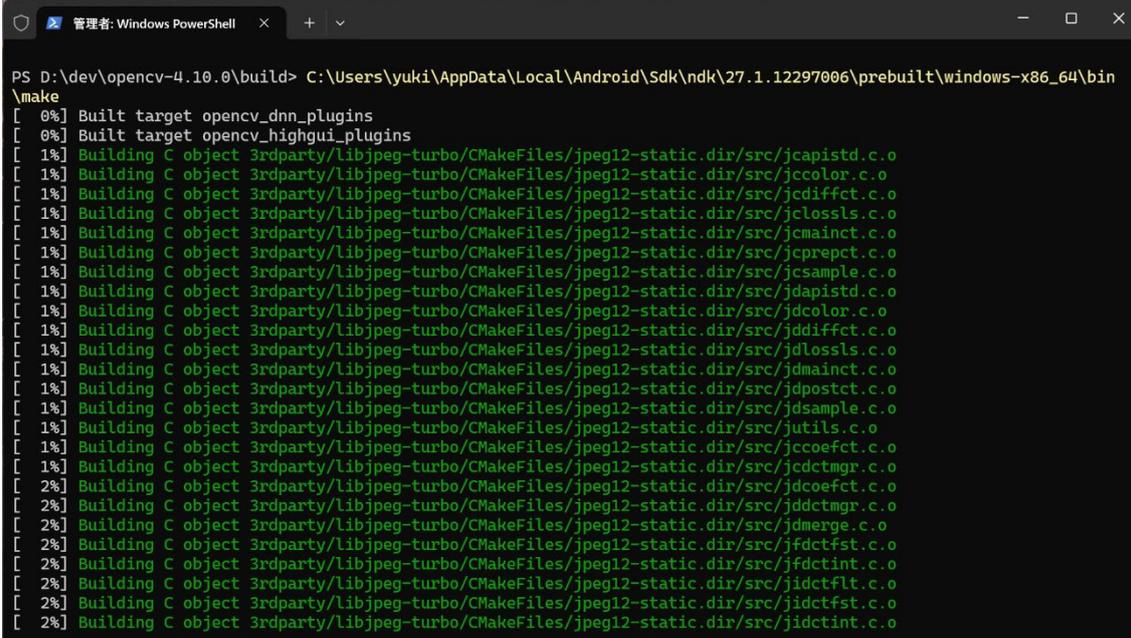

Open a terminal and move the current directory to "D:¥dev-4.10.0¥opencv-4.10.0¥build", then type "C:¥Users¥yuki¥AppData¥Local¥Android¥Sdk¥ndk¥27.1.12297006¥prebuilt¥windows-x86_64¥bin¥make" and execute it. When the process is finished, the next step is to type "C:¥Users¥yuki¥AppData¥Local¥Android¥Sdk¥ndk¥27.1.12297006¥prebuilt¥windows-x86_64¥bin¥make install" and execute it. Then the built binary will appear in "D:¥dev¥opencv-4.10.0¥build".

If you want to run architectures other than Arm64-v8a, you can rewrite the CMAKE GUI option setting ANDROID_ABI:STRING from arm64-v8a to other architectures such as armeabi-v7a or x86_64 and build again.

## Use the built OpenCV in Android Studio

To use the built OpenCV in your Android Studio project, open Android Studio, go to "New Project" and create a "Native C++" project.

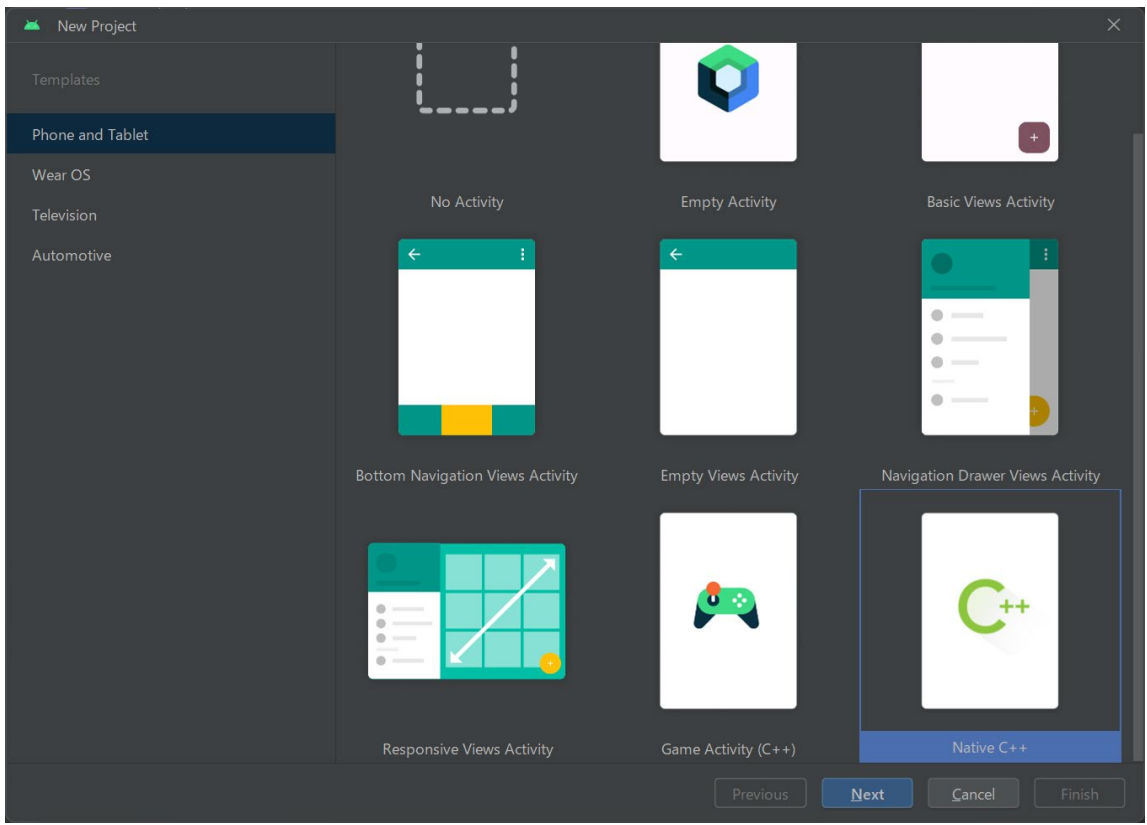

After the project is created, select "File" > "New" > "import Module".

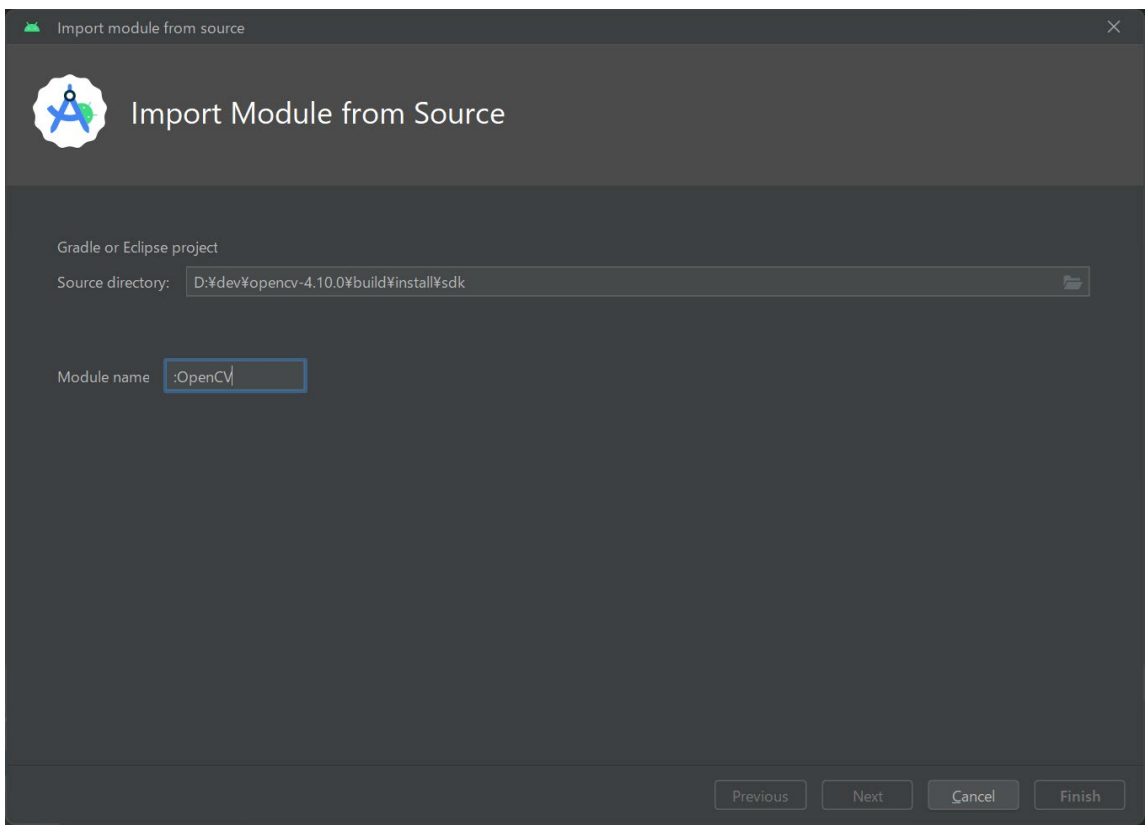

Enter "D:¥dev¥opencv-4.10.0¥build¥install¥sdk" in the Source directory and ":OpenCV" in the Module name. Click 'Finish'.

Next, select "File" and then "Project Structure".

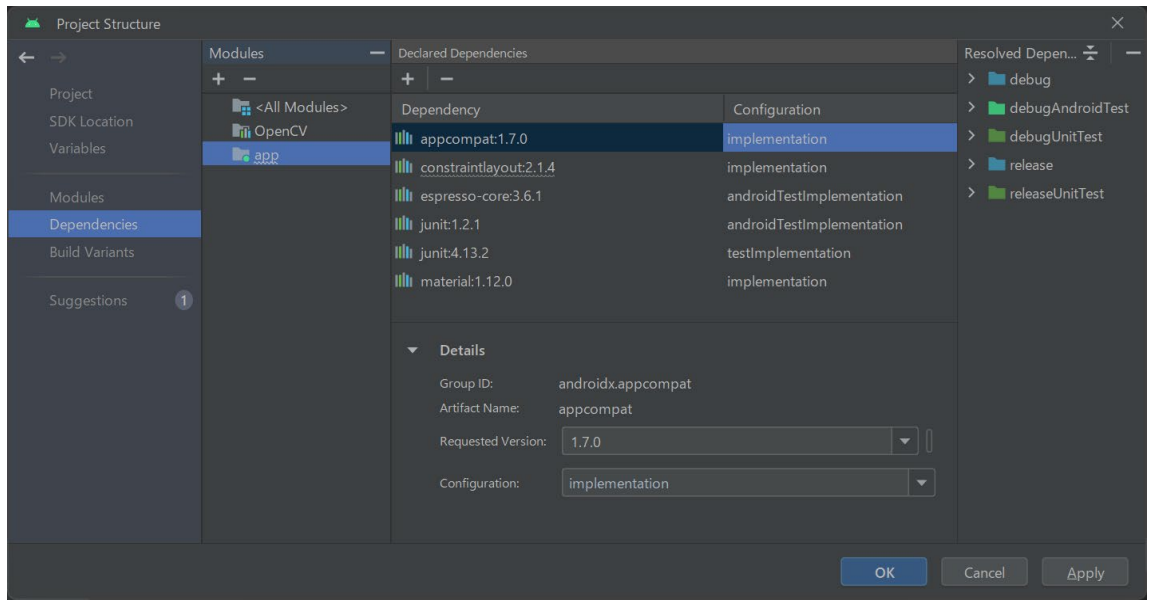

Select "Dependencies", then "app", and click "+" under "Declared Dependencies".

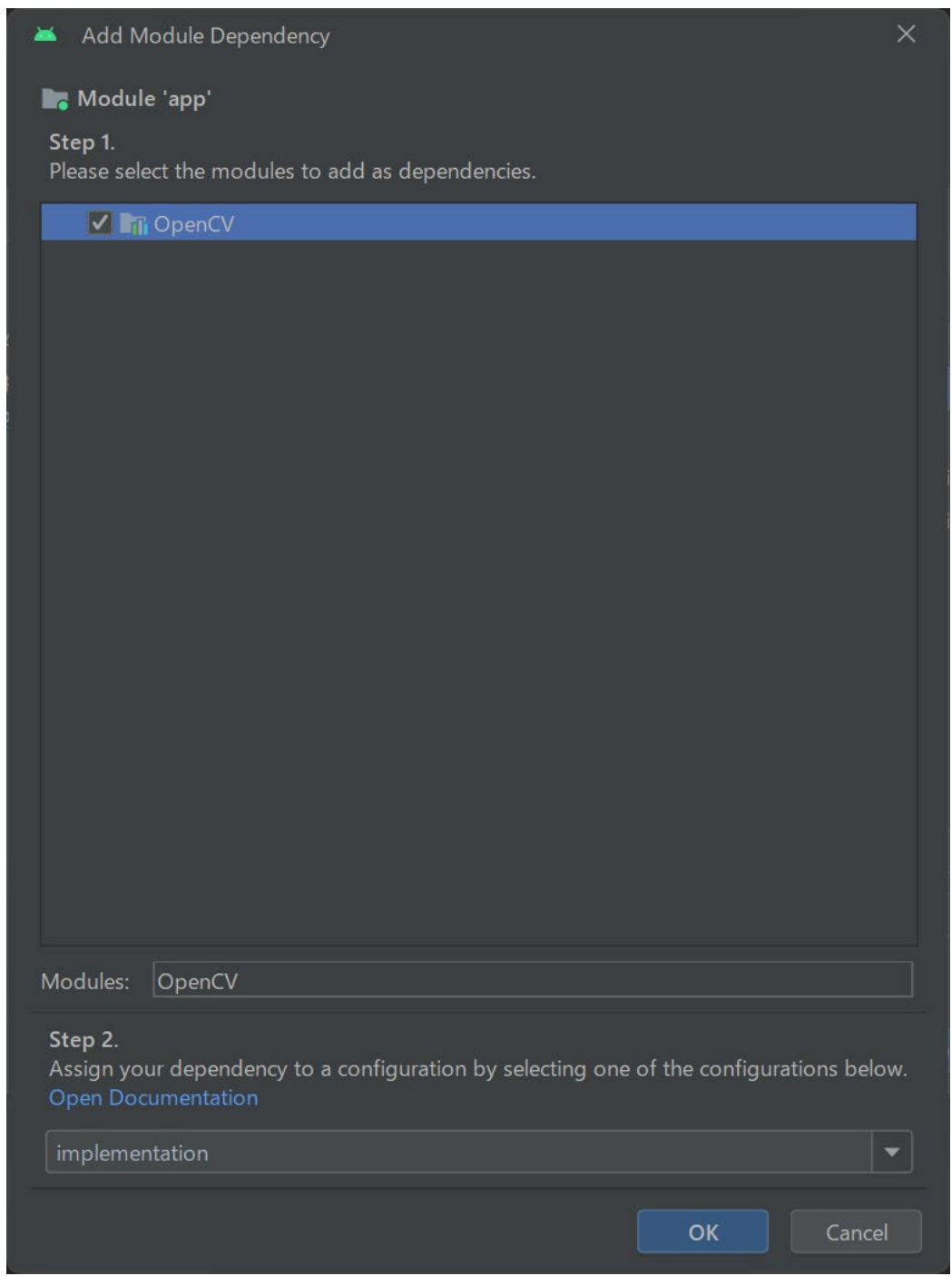

Check "OpenCV" and click 'OK'.

Comment out the "apply plugin: 'kotlin-android'" section of :OpenCV's build.gradle and add the following code.

```
    buildFeatures{
        aidl true
        buildConfig true
    }

    namespace 'org.opencv'//Add namespace
```

Add the following code to AndroidManifest.xml of app.

```
<uses-native-library
    android:name="libOpenCL.so"
    android:required="false"/>
```

Add the following code to CMakeLists.txt in app.

```
set( OPENCV_INCLUDE_DIR "${PROJECT_SOURCE_DIR}/../../../../OpenCV/native/jni/include")
set( OPENCV_LIB_DIR "${PROJECT_SOURCE_DIR}/../../../../OpenCV/native/libs" )
include_directories(${OPENCV_INCLUDE_DIR})
add_library( lib_opencv SHARED IMPORTED )
set_target_properties(lib_opencv PROPERTIES IMPORTED_LOCATION
${OPENCV_LIB_DIR}/${ANDROID_ABI}/libopencv_java4.so)

target_link_libraries(${CMAKE_PROJECT_NAME}
        lib_opencv
        android
        log)
```

For information on how to write GPGPU source code using OpenCL, please refer to the sample project "https://github.com/cardinal-casket-yuki-n/CLSample" which performs angular spectrum processing on images captured by a camera. (Pre-built OpenCV is not included in the project.)